\documentclass{JHEP3}

\def\be{\begin{equation}}
\def\ee{\end{equation}}
\def\bear{\begin{eqnarray}}
\def\eear{\end{eqnarray}}
\def\nn{\nonumber}

\def\half{{{1\over 2}}}

\newcommand\inv[1]{{1\over{#1}}}

\def\a{{\alpha}}

\def\e{{\epsilon}}

\preprint{PUPT-2014 \\ LPTENS-0137 \\ \tt {hep-th/0111133}}

\author{ Joanna L. Karczmarek and and Curtis G. Callan, Jr. \\
Department of Physics, Jadwin Hall \\ 
Princeton University \\ 
Princeton, NJ 08544, USA \\
\email{joannak@princeton.edu, callan@feynman.princeton.edu}}

\title{Tilting the Noncommutative Bion}

\abstract{ 
In this note, we extend the noncommutative bion core solution 
of Constable, Myers and Tafjord \cite{ncbion} to include the effects
of a nonzero NS-NS two-form $B$. The result is a `tilted bion',
in which the core expands out to a single D3-brane at an angle to the 
D1-brane core. Its properties agree perfectly with an analysis of the 
dual situation, that of a magnetic charge on an abelian D3-brane in a 
background worldvolume magnetic field. We also demonstrate that this 
agreement extends beyond geometry to include the field strength on 
the D3-brane.  We make a proposal for including possible worldvolume 
gauge fields when mapping a noncommutative geometrical brane solution 
onto a corresponding commutative brane description.} 


\begin{document}

\section{Introduction}\label{intro}
Much effort has been devoted to generalizing the Born--Infeld action so
as to describe the dynamics of multiple superposed D-branes. It is known that
the gauge group for a stack of $N$ superposed D-branes is enhanced 
from $U(1)^N$ to $U(N)$ and that the brane worldvolume supports a $U(N)$  
gauge field as well as a set of scalars in the adjoint representation
of $U(N)$ (one for each of the transverse coordinates). A specific
proposal for a generalized action involving such fields has recently 
been given by Myers in \cite{dielectric} (see also references in
this paper for other work on this problem). For a Dp-brane in static
gauge, Myers' action takes the following, rather forbidding, form: 
\bear 
\label{eqn:SBI} 
&~& S_{BI} = -T_p \int d^{p+1}\sigma STr  
\\ 
&~& \left( e^{-\phi} \sqrt{ 
-\det (P_{ab}[(G+\lambda B)_{\mu\nu}+(G+\lambda B)_{\mu i}(Q^{-1}-\delta)^{ij}
 (G+\lambda B)_{j\nu}] + \lambda {\mathcal{F}}_{ab}) det(Q^i_j) }\right) ~,
\nn 
\eear 
where 
\be 
\label{eqn:Q}
Q^i_j \equiv \delta^i_j + i\lambda[\Phi^i,\Phi^k](G+\lambda B)_{kj}~,
\ee 
$\lambda \equiv 2 \pi \a'$, $a,b$ are indices in the worldvolume of the 
Dp-branes, $i,j,k$ are in the transverse space and $\mu,\nu$ are 
ten-dimensional indices. The $\Phi^i$ are $N \times N$ matrix scalar 
fields describing the transverse displacements of the branes and 
${\mathcal{F}}_{ab}$ is the worldvolume gauge field (both fields are
in the adjoint representation of $U(N)$). The symbol $P_{ab}[M]$
stands for a pullback of the ten-dimensional matrix $M$ to the brane 
worldvolume in which the matrix coordinates $\Phi^i$ define the surface to 
which to pull back and all derivatives of these coordinates are made
gauge covariant. Finally, $STr$ is Tseytlin's symmetrized trace 
operation \cite{symtrace}. We refer the reader to Myers' paper 
\cite{dielectric} for a more complete definition and a discussion 
of useful simplifying approximations. 
Whether or not this action is ``exact'', it does seem 
to capture a lot of information about such structures as the commutators 
of the $\Phi$'s with themselves, which vanish in the $U(1)$ limit and 
cannot be directly inferred from the abelian Born-Infeld action.
 
In \cite{ncbion}, Myers' action was applied to a stack of $N$ D1-branes 
in a flat background and it was shown that the transverse coordinates 
of the D1-branes `flare-out' to a flat three-dimensional space.  
This was interpreted as a description of a collection of D1-branes attached to
an orthogonal D3-brane, one of the standard D-brane intersections. 
This situation has a `dual' description in terms of a single (abelian) 
D3-brane carrying a point magnetic charge. Using the Born-Infeld
action, one finds that a magnetic monopole of strength $N$ produces 
a singularity in the D3-brane's transverse displacement, a `spike' 
which can be interpreted as $N$ D1-strings attached to the D3-brane 
\cite{callan,gibbons}. To the extent that it is possible to compare them,
there is a perfect quantitative match between the two pictures.
The states in question are BPS states and this match is evidence 
that the Myers' action (\ref{eqn:SBI}) captures the dynamics 
of BPS states at least. 
 
In this brief note, we extend these considerations to a more
complex example of a BPS state: we generalize the solution of 
\cite{ncbion} to include a nonzero background NS-NS two-form field B, 
in a direction transverse to the D1-branes, but 
parallel to the emergent D3-brane.  In the `dual'  
description on the D3-brane, the B field becomes a $U(1)$ magnetic
field on the D3-brane which pulls on the magnetic charge
associated with the D1-brane. Simple force balance considerations
suggest that the spike representing the D1-branes should be tilted 
away from the direction normal to the D3-brane, and this is precisely
what is found in explicit solutions of the Born-Infeld equations  
\cite{hashimoto, moriyama}. We will show that this system can be
analysed in the nonabelian D1-brane picture and that
there is, once again, a perfect quantitative match between the
results of the two dual calculations. One slight novelty of our 
approach is that we can demonstrate this agreement not only for
geometrical quantities, but also for worldvolume gauge fields.
 
The paper is organized as follows:
In section \ref{sec-D3}, we review how the bion arises 
in the abelian theory on a D3-brane, both with 
and without a B-field. In section \ref{sec-D1}, we review 
the construction of the nonabelian bion, and 
then extend it to a nonzero B-field. In section \ref{sec-plane}, 
we take a brief detour and describe how lower-dimensional D-branes 
can form flat, noncommutative, higher dimensional D-branes equipped
with worldvolume gauge fields. In this context we show how the
higher-dimensional worldvolume gauge field is constructed out of
the lower-dimensional noncommuting coordinates. Finally, in section 
\ref{sec-curved}, we apply this recipe to the gauge field
on the bion of section \ref{sec-D1} and show that 
there is perfect agreement between the two approaches, in both
the geometry and the field strength on the brane.


\section{The Bion solution on a D3-brane} 
\label{sec-D3} 
Consider the abelian Born-Infeld action for a single 
D3-brane in flat space. Let the D3-brane extend in 0123-directions, 
and let the coordinates on the brane be denoted by $x^i$, 
$i=0,\ldots,3$. Restricting the brane to have displacement in
only one of the transverse directions, we can take the (static gauge)
embedding coordinates of the brane in the ten-dimensional space to be  
$X^i = x^i$, $i=0,\ldots,3$; $X^a = 0$, $a=4,\ldots,8$;
$X^9 = \sigma(x^i)$. Then there exists a static (BPS) solution of 
the Born-Infeld action corresponding to placing $N$ units of
$U(1)$ magnetic charge at the origin of coordinates on the brane 
\cite{callan}: 
\be 
X^9(x^i) = \sigma(x^i) = {{q} \over {\sqrt{(x^1)^2+(x^2)^2+(x^3)^2}}} ~,
\label{eqn:bion0} 
\ee 
where $q=\pi\alpha' N$, and $N$ is an integer. This magnetic
bion solution to the Born--Infeld action corresponds to $N$ superposed
D1-strings attached to the D3-brane at the origin. It is ``reliable''
in the sense that the effect of unknown higher-order corrections in 
$\alpha'$ and $g$ to the action can be made systematically small
in the large-$N$ limit (see \cite{callan} for details). 
At a fixed $X^9=\sigma$, the cross-section of the deformed
D3-brane is a 2-sphere with a radius  
\be 
r(\sigma) = {{ \pi\alpha' N} \over {\sigma}}~.
\label{eqn:r0} 
\ee 
 
In the presence of a two-form field B parallel to the world volume of 
the brane,  ($\half B dx^1\wedge dx^2$ to be concrete), the above solution
is modified as follows  \cite{moriyama}:
\be 
\label{eqn:bionB}
{\sigma \over \cos(\alpha)}=
{{q} \over {\sqrt{(x^1)^2+(x^2)^2+\cos(\alpha)^2 
(x^3-\tan(\alpha)\sigma)^2}}}~. 
\ee 
where $\tan(\alpha) =2\pi\alpha' B$. Because the transverse displacement
$\sigma$ is not a single-valued function of the base coordinates $x^i$,
the geometry is somewhat obscure. It is more transparent 
in rotated coordinates defined by
\be
\label{eqn:y-coord}
Y^1=X^1, \quad Y^2=X^2, \quad Y^3=\cos(\a) X^3 - \sin(\a) X^9, \quad
Y^4=\sin(\a) X^3 + \cos(\a) X^9. 
\ee
Choosing $Y^{1,2,3}$ as the worldvolume coordinates, the embedding becomes
\be
\label{eqn:bionY}
Y^{(1,2,3)}(y) = y^{(1,2,3)}, \quad Y^4(y) = \tan(\a)~ y^3 + {q\over{
\sqrt{(y^1)^2+(y^2)^2+(y^3)^2}}}~.
\ee
It is easy to see that this describes D1-strings tilted away from 
the normal to the D3-brane by an angle $\alpha$ (when $B\to 0$,
$\alpha\to 0$, and the branes become orthogonal).

Reverting to the original coordinates (\ref{eqn:bionB}),
we can show that the D3-brane at a fixed transverse displacement 
$X^9=\sigma$ is an ellipsoid of revolution defined by the equation
\be
1 = {x_1^2\over r_1^2(\sigma)}+{x_2^2\over r_2^2(\sigma)}
+{(x_3-\sigma \tan (\alpha))^2\over r_3^2(\sigma)}
\ee
with major and minor axes
\be \label{eqn:rB}
r_1(\sigma)=r_2(\sigma)=\cos(\alpha) {{\pi\alpha' N } \over {\sigma}} ~,\qquad
r_3(\sigma)= {{\pi\alpha' N} \over {\sigma}}~.
\ee 
For large $\sigma$, the ellipsoid becomes small and defines a slice
through the D1-brane that is attached to the D3-brane. The ellipsoid
is centered at brane coordinates $(X_1,X_2,X_3)=(0,0,\tan(\alpha)\sigma)$ 
and the fact that $X_3$ varies linearly with $\sigma$ implements the
tilt of the D1-brane. The tilting arises for simple reasons of 
force balance. The D1-brane spike 
behaves like a magnetic charge from the point of view of the
worldvolume gauge theory; the background B field is equivalent
to a uniform magnetic field on the D3-brane and exerts a force on
the magnetic charge which must be balanced by a component of the 
D1-brane tension along the D3-brane . 

The gauge field on the brane is particularly easy to obtain
in the tilted coordinates (\ref{eqn:y-coord}). The configurations
under discussion are not the most general solution of the equations
of motion: they are special minimal energy solutions that satisfy
the BPS condition (and preserve some supersymmetry). The BPS condition 
relates the total 2-form field on the brane to the divergence of the 
transverse displacement scalar as follows \cite{callan,gibbons}:
\be
(2\pi\a') \epsilon^{ijk} (F + B)_{jk} = \pm
{\partial \over {\partial y^i}} Y^4
= {\mp q\over [(y^1)^2+(y^2)^2+(y^3)^2]^{3/2}}~ y^i + \tan(\a) \delta^i_3~.
\ee
The ambiguous sign encodes the difference between a D3- and a
$\overline {\textrm{D3}}$-brane. From this we read off that
the magnetic field ${\mathcal B}_k \equiv \epsilon^{ijk} F_{jk}$ is just
\be
\label{eqn:Ffield}
{\mathcal B}_i(y) = \mp{N\over 2 [(y^1)^2+(y^2)^2+(y^3)^2]^{3/2}}~ y^i~,
\ee
or a Coulomb field due to $N$ charges. The end of the D1-brane(s) acts as a 
magnetic charge and the space-time $B$ field provides an effective
uniform magnetic field which exerts a force on the magnetic charge,
hence tilting the D1-branes.

Note that for the upper sign (D3-brane), when N is positive, the 
D1-brane(s) run `towards' the D3-brane, and when it is negative, 
they run `away' from the D3-brane. For the lower sign ($\overline 
{\textrm{D3}}$-brane case), the end of the 
$\overline {\textrm{D1}}$-brane(s) running 
`towards' the $\overline {\textrm{D3}}$-brane acts as a positive 
magnetic source and the end of the $\overline {\textrm{D1}}$-brane(s) 
running `away' from $\overline {\textrm{D3}}$-brane acts as negative one.
We will encounter the same four cases in our dual treatment by the
nonabelian D1-branes.

\section{Dual treatment by nonabelian D1-branes}
\label{sec-D1} 
In describing the intersection of one D3-brane with $N$ D1-branes,
one has the option of starting from the dynamics of the D3-brane
and trying to derive the D1-branes (this was the approach of the
previous section), or of starting from the nonabelian dynamics of 
multiple D1-branes and trying to derive the D3-brane. The latter 
approach has been applied in \cite{ncbion} to the case in which there
is no background $B$ field. In this section we will review that work
and show how to generalize it to the case where $B\ne 0$ and the
bion is tilted.

We begin by reviewing the results from \cite{ncbion} on the $B=0$ case, 
while introducing some notation which will be useful later.
We consider the nonabelian Born--Infeld action of 
equation (\ref{eqn:SBI}) specialized to the case of N coincident D1-branes,
flat background spacetime ($G_{\mu\nu}=\eta_{\mu\nu}$), vanishing
$B$ field, vanishing worldvolume gauge field and constant dilaton. 
The action then depends only on the $N\times N$ matrix transverse 
scalar fields $\Phi^i$'s. In general, $i=1,\ldots,8$, but since we are
interested in studying the D1/D3-brane intersection, we will allow
only three transverse coordinate fields to be active ($i=1,2,3$). 
The explicit reduction of the static gauge action ($X^0 = \tau$ and 
$X^9=\sigma$) is then
\be
\label{eqn:SBI1}
S_{BI} = -T_1 \int d\sigma d\tau STr 
 \sqrt{ -\det(\eta_{ab}+\lambda^2\partial_a\Phi^i 
   Q^{-1}_{ij}\partial_b\Phi^j) det(Q^{ij}) } ~,
\ee 
where
\be 
Q^{ij} = \delta^{ij} + i\lambda[\Phi^i,\Phi^j]~.
\ee 
Since the dilaton is constant, we incorporate it in the 
tension $T_1$ as a factor of $g^{-1}$. 
 
We look for static solutions ({\it{i.e.}}, $\Phi = \Phi(\sigma)$
only). Since we have no hope of finding a general static solution of 
these nonlinear matrix equations, we make some simplifying 
assumptions which have a chance of being valid on the restricted 
class of BPS solutions. The action (\ref{eqn:SBI1}) depends only 
on the two matrix structures $\partial_a\Phi^i$ and
$W_i \equiv \half i \epsilon_{ijk}[\Phi^j,\Phi^k]$ and, because of 
the nature of the $STr$ instruction, they may be treated as commuting 
quantities until the final step of doing the gauge trace to evaluate
the action. This allows us to evaluate the determinants in the 
definition of the action (\ref{eqn:SBI1}) and to convert the energy
functional to the following form:
\be 
U_{B=0} = \int d \sigma STr \sqrt {1 + \lambda^2(\partial \Phi^i)^2 
+ \lambda^2(W_i)^2 + \lambda^4 (\partial \Phi^i W_i)^2 } ~.
\label{eqn:S0} 
\ee 
Continuing to treat $\Phi^i$ and $W^i$ as commuting objects, we see that
this energy functional can be written as a sum of two squares
\be 
U_{B=0} = \int d \sigma STr \sqrt { 
(1 \pm \lambda^2 \partial \Phi^i W_i)^2 
+ \lambda^2 (\partial \Phi^i \mp W_i)^2} ~,
\ee 
and is minimized by a displacement field satisfying the 
first-order BPS-like equation $\partial \Phi^i = \pm W_i$. 
This equation, written more explicitly as
\be 
\partial \Phi_i = \pm \half i \epsilon_{ijk}[\Phi^j,\Phi^k] ~,
\label{eqn:BPS0} 
\ee 
is known as the Nahm equation \cite{nahm0}. The $\pm$ ultimately 
corresponds to the choice between a bundle of D1- or 
$\overline {\textrm{D1}}$-branes. The Nahm equation is a very 
plausible candidate for the exact equation to be satisfied by a BPS 
solution of this system and the fact that the Myers action 
(\ref{eqn:SBI1}) implies it in the BPS limit is very satisfactory.  

The Nahm equation has the trivial solution $\Phi=0$ which
corresponds to an infinitely long bundle of coincident D1-branes. 
In \cite{ncbion}, a much more interesting solution was found
by starting with the following ansatz: 
\be 
\Phi^i = \hat R(\sigma) \alpha^i,\qquad\qquad 
(\alpha^1, \alpha^2, \alpha^3) \equiv {\bf{X}} ~,
\label{eqn:ansatz0} 
\ee 
where $\alpha^i$ form an $N\times N$ representation of the 
generators of an $SU(2)$ subgroup of $U(N)$, 
$[\alpha^i, \alpha^j] = 2 i\epsilon_{ijk}\alpha^k$. 
With this ansatz, both $\partial\Phi^i$ and $W^i$ are proportional
to the generator matrix $\alpha^i$.
When the ansatz is substituted into the BPS condition
(\ref{eqn:BPS0}), we obtain a simple equation for $\hat R$,  
\be 
\hat R'= \mp 2\hat R^2 ~,
\ee 
which is solved by 
\be 
\hat R = \pm \inv{2 \sigma}~.
\label{eqn:R0} 
\ee 
Substituting the ansatz (\ref{eqn:ansatz0}) into  
(\ref{eqn:S0}) leads to the following effective action
for $\hat R(\sigma)$ :
\be 
U_{B=0}[\hat R(\sigma)] = \int d \sigma STr \sqrt { 
(1 + \lambda^2(\hat R')^2 {\bf{X}}^2) 
(1 + 4\lambda^2(\hat R)^4 {\bf{X}}^2) 
}~. 
\label{eqn:S(R)0} 
\ee 
It can be shown that (\ref{eqn:R0}) satisfies the 
equations of motion following from the action (\ref{eqn:S(R)0}). 

This solution maps very nicely onto the bion solution of the 
previous section.  At a fixed point $|\sigma|$ on the D1-brane stack,
the geometry given by (\ref{eqn:ansatz0}) is that of a sphere
with the physical radius $R^2={\lambda^2\over N}Tr (\Phi^i)^2$
(the only sensible way to pass from the matrix transverse displacement
field $\lambda\Phi^i$ to a pure number describing the geometry).
For the ansatz under consideration, this gives
\be 
R(\sigma)^2 = {{\lambda^2}\over {N}} Tr (\Phi^i)^2 
= \lambda^2 \hat R(\sigma)^2 C ~,
\ee 
where $C$ is the quadratic Casimir, equal to $N^2-1$ 
for an irreducible representation of $SU(2)$. This gives 
\be 
\label{eqn:R}
R(\sigma)= {{\lambda \sqrt{N^2-1}} \over {(2 |\sigma|)}} 
\cong {{\pi\alpha' N} \over {|\sigma|}} 
\ee 
for large N, in agreement with equation (\ref{eqn:r0}). This 
completes our synopsis of the arguments given in \cite{ncbion} for
the agreement between the commutative and noncommutative approaches
to the D1/D3-brane intersection.

A simple argument can be made at this point to strengthten the
meaning of equation (\ref{eqn:R}).  At a fixed $\sigma$, the 
Fourier transform of the density of the $D1$-strings is given
by \cite{mark}
\be
\label{density}
\widetilde \rho (k) = Tr \left ( e^{i\lambda k_i \Phi^i} 
\right )~.
\ee
This is simply the operator to which the $09$-component
of the RR 2-form $C^{(2)}$, $C_{09}$, couples.
For the solution (\ref{eqn:ansatz0}), this is evaluated to
give
\be
\widetilde \rho (k) = {\sin(\lambda N \hat R |k|)
\over \sin(\lambda \hat R |k|)}~,
\ee
which for $N\gg 1$, and $k$ such that 
$(\lambda N \hat R)^{-1} < |k| \ll (\lambda \hat R)^{-1}$
(i.e., for momentum large enough to resolve the size of the
sphere, but not large enough to resolve the individual brane 
constituents) gives
\be
\widetilde \rho (k) \cong {\sin(\lambda N \hat R |k|)
\over \lambda \hat R |k|}~.
\ee
This is precisely the Fourier transform of the density
distribution representing a thin shell,
\be
\widetilde \rho (k) = 
\int d^3 x e^{i\vec k\cdot\vec x}\rho(x) \quad \mbox{for}
\quad \rho(x) = {N \over 4 \pi R^2} \delta(|x|-R)
\quad \mbox{with} \quad R = \lambda N \hat R = 
{\pi \a' N \over |\sigma|}~,
\ee
in agreement with equation (\ref{eqn:R}).

To distinguish the various cases involving branes and antibranes, note
that $\sigma$ can either run from $-\infty$ to $0$, in which case
the D1($\overline {\textrm{D1}}$)-branes run `towards' the 
D3($\overline {\textrm{D3}}$)-brane plane, or from $0$ to $\infty$, 
in which case the D1($\overline {\textrm{D1}}$)-branes run `away' 
the D3($\overline {\textrm{D3}}$)-brane plane.  We have thus four cases:
\begin{itemize}
\item[{\bf{A}}] :~
Stack of D1-branes expanding to a D3-brane
, $\partial \Phi^i = + W_i$ and $\sigma \in (-\infty, 0)$
\item[{\bf{B}}] :~
Stack of D1-branes expanding to a D3-brane
, $\partial \Phi^i = + W_i$ and $\sigma \in (0, \infty)$
\item[{\bf{C}}] :~
Stack of $\overline {\textrm{D1}}$-branes expanding to a $\overline {\textrm{D3}}$-brane
, $\partial \Phi^i = - W_i$ and $\sigma \in (-\infty, 0)$
\item[{\bf{D}}] :~
Stack of $\overline {\textrm{D1}}$-branes expanding to a $\overline {\textrm{D3}}$-brane
, $\partial \Phi^i = - W_i$ and $\sigma \in (0, \infty)$
\end{itemize}
\begin{figure}

\begin{center}
\setlength{\unitlength}{0.75in}
\begin{picture}(2,0)

\put(-3.3,-1.3){{\bf{A}}}
\put(-3.5,-1){\line(2,1){1}}
\put(-3.5,-1){\line(2,0){1.5}}
\put(-2.0,-1){\line(2,1){1}}
\put(-2.5,-0.5){\line(2,0){1.5}}
\put(-1.8,0.25){\vector(-1,-2){0.25}}
\put(-2.05,-0.25){\line(-1,-2){0.25}}
\put(-2.3,-0.75){\circle*{0.05}}
\multiput(-2.3,-0.75)(0,0.1){6}{\line(0,1){0.05}}
\put(-2.25,-0.4){$\a$}
\put(-2.39,-0.92){$+$}
\put(-1.75,-.7){\small{D3}}
\put(-1.8,0){\small{D1}}

\put(-1.3,-1.3){{\bf{B}}}
\put(-1.5,-1){\line(2,1){1}}
\put(-1.5,-1){\line(2,0){1.5}}
\put(0.0,-1){\line(2,1){1}}
\put(-0.5,-0.5){\line(2,0){1.5}}
\put(-0.25,-0.75){\vector(-1,-2){0.25}}
\put(-0.5,-1.25){\line(-1,-2){0.25}}
\put(-0.25,-0.75){\circle*{0.05}}
\multiput(-0.25,-0.75)(0,-0.1){6}{\line(0,-1){0.05}}
\put(-0.4,-1.2){$\a$}
\put(-0.31,-0.67){$-$}
\put(0.25,-.7){\small{D3}}
\put(-.6,-1.7){\small{D1}}

\put(0.7,-1.3){{\bf{C}}}
\put(0.5,-1){\line(2,1){1}}
\put(0.5,-1){\line(1,0){1.5}}
\put(2.0,-1){\line(2,1){1}}
\put(1.5,-0.5){\line(1,0){1.5}}
\put(1.2,0.25){\vector(1,-2){0.25}}
\put(1.45,-0.25){\line(1,-2){0.25}}
\put(1.7,-0.75){\circle*{0.05}}
\multiput(1.7,-0.75)(0,0.1){6}{\line(0,1){0.05}}
\put(1.55,-0.4){$\a$}
\put(1.62,-0.92){$-$}
\put(2.3,-.7){$\overline{\mbox{\small{D3}}}$}
\put(1.35,0){$\overline{\mbox{\small{D1}}}$}

\put(2.7,-1.3){{\bf{D}}}
\put(2.5,-1){\line(2,1){1}}
\put(2.5,-1){\line(1,0){1.5}}
\put(4,-1){\line(2,1){1}}
\put(3.5,-0.5){\line(1,0){1.5}}
\put(3.75,-0.75){\vector(1,-2){0.25}}
\put(4.0,-1.25){\line(1,-2){0.25}}
\put(3.75,-0.75){\circle*{0.05}}
\multiput(3.75,-0.75)(0,-0.1){6}{\line(0,-1){0.05}}
\put(3.8,-1.2){$\a$}
\put(3.69,-0.67){$+$}
\put(4.25,-.7){$\overline{\mbox{\small{D3}}}$}
\put(3.85,-1.7){$\overline{\mbox{\small{D1}}}$}

\end{picture}
\end{center}
\vspace{1in}
\caption{The four cases {\bf{A}}-{\bf{D}} discussed in the text.}
\label{figure}
\end{figure}
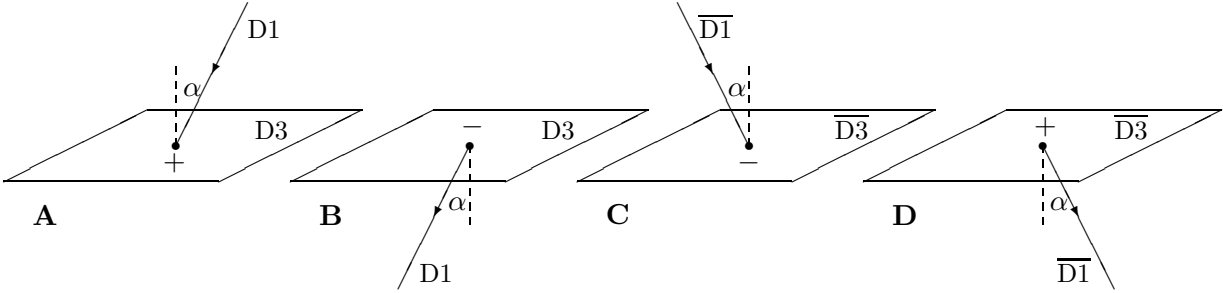
Cases {\bf{A}} and {\bf{D}} correspond to the D1($\overline {\textrm{D1}}$)-branes 
running `towards' (`away from') the D3($\overline {\textrm{D3}}$)-brane plane, 
and thus should represent a positive magnetic charge, while
Cases {\bf{B}} and {\bf{C}} should represent a negative magnetic charge.
We will see shortly that this is the case and that the four cases
match the four cases found in the abelian treatment of the same
problem starting from the D3-brane (see the end of section \ref{sec-D3}).

Our next step is to turn on the background $B$ field in order to study 
the tilted bion from the noncommutative D1-brane point of view.
The only difference from the previous case is that we turn on the component
$B_{12}=\mbox{const.}$ of the background $B$ field (remember that the
D1-brane worldvolume spans the $(X^0,X^9)$ plane and that only the
$i=1,2,3$ components of the matrix transverse displacement field are
allowed to be nonzero). It is still the case that the action is
a functional only of the fields $\partial\Phi^i$ and  
$\epsilon_{ijk}[\Phi^j,\Phi^k]$ and the same reasoning as
before leads us to treat these quantities as commuting objects 
inside the $STr$ instruction. In this way, the action (\ref{eqn:SBI}) 
can be reduced to something much more explicit. To get the most 
transparent results, it helps to define rescaled fields 
\be 
\varphi_1 = \sqrt{(1+\lambda^2 B^2)} \Phi_1~, \qquad 
\varphi_2 = \sqrt{(1+\lambda^2 B^2)} \Phi_2~, \qquad 
\varphi_3 = \Phi_3~,
\label{eqn:rescale} 
\ee 
and to redefine the commutator $W^i$ as 
\be 
W_i \equiv \half i \epsilon_{ijk}[\varphi^j,\varphi^k] -
\delta^3_i B ~.
\ee 

After some rather tedious algebra (made much easier by MAPLE) 
to evaluate the determinants in the definition of the action,
we get a result for the energy functional that is almost identical
to (\ref{eqn:S0}):
\bear 
U_{B\neq 0} &=& {{1}\over{\sqrt{1+\lambda^2 B^2}}} 
\int d \sigma STr \sqrt{ 
1 + \lambda^2(\partial \varphi^i)^2 
+ \lambda^2(W_i)^2 
+ \lambda^4 (\partial \varphi^i W_i)^2 
} \nn \\ &=& 
{{1}\over{\sqrt{1+\lambda^2 B^2}}} 
\int d \sigma STr \sqrt{ 
(1 \pm \lambda^2 \partial \varphi^i W_i)^2 
+ \lambda^2 (\partial \varphi^i \mp W_i)^2 
} ~.
\label{eqn:SB} 
\eear 
The action is still `linearized' by taking 
$\partial \varphi^i = \pm W_i$, which means that the
BPS condition in the presence of a background $B$ field is
\be 
\partial \varphi_i = \pm i (\half  
\epsilon_{ijk}[\varphi^j,\varphi^k] + \delta^3_i i B)~.
\label{eqn:BPSB} 
\ee 
This is precisely the generalization of the Nahm equation 
that has been derived in the context of studies of magnetic 
monopoles in noncommutative field theory \cite{nahmB}. It is  
a plausible candidate for the exact BPS condition for the nonabelian
D1-brane system and we will show that it gives a detailed account 
of the physics of the tilted bion. The fact that the generalized 
Nahm equation is implied by the Myers action is further evidence 
for the essential correctness of the latter.

In order to solve the modified equations of motion, we have to
slightly modify the ansatz (\ref{eqn:ansatz0}), expressing the
fields $\varphi$ in terms of generators of an N-dimensional 
representation of $SU(2)$ and a scalar function $\hat R(\sigma)$:
\be 
\varphi^i = \hat R(\sigma) \alpha^i - \delta^3_i 
{{B} \over {2 \hat R(\sigma)}}~ .
\label{eqn:ansatzB} 
\ee 
When this modified ansatz is substituted into the BPS 
equation (\ref{eqn:BPSB}), we obtain the same equation for 
$\hat R$ as before, namely $\hat R'= \mp 2\hat R^2$: 
solution (\ref{eqn:R0}) still holds. If we collect the generators
into a modified triplet ${\bf{X}}\equiv
(\alpha^1, \alpha^2, \alpha^3+ {{B}\over{2{\hat R}^2}})$ and use
(\ref{eqn:ansatzB}), the action (\ref{eqn:SB}) can be expressed 
as an effective action for $\hat R(\sigma)$:
\be 
U_{B\ne 0}[\hat R(\sigma)] =  
{{1}\over{\sqrt{1+\lambda^2 B^2}}} 
\int d \sigma STr \sqrt { 
(1 + \lambda^2(\hat R')^2 {\bf{X}}^2) 
(1 + 4\lambda^2(\hat R)^4 {\bf{X}}^2) } ~.
\label{eqn:S(R)B} 
\ee 
This looks the same as (\ref{eqn:S(R)0}) but is not quite 
because ${\bf{X}}$ now depends on $\hat R$. Nevertheless,
the same radial function (\ref{eqn:R0}) continues to be a solution,
further testing the compatibility of the action with the BPS 
condition.  Notice that this does not conclusively prove that 
the ansatz (\ref{eqn:ansatzB}) with (\ref{eqn:R0}) is a solution to 
the full equations of motion implied by (\ref{eqn:SB}). 
 
It is easily seen that this solution corresponds to the 
tilted bion solution discussed in the previous section. 
Equation (\ref{eqn:rescale}) matches the ratios of axes
given in (\ref{eqn:rB}). The over-all size of the spheroid 
agrees with (\ref{eqn:rB}) by an argument identical to that 
given for $B=0$. The shift of its center is given by 
\be
\Delta^i(\sigma)= {1\over N}tr(\lambda\Phi^i)=\Delta \delta^i_3
\ee
(by virtue of the fact that $tr(\alpha^i)=0$), where
\be
\Delta = - {\lambda {B}\over{2\hat R}} = \mp \lambda B \sigma = 
\left \{ \begin{array}{ll}
\tan(\alpha) |\sigma| &\textrm{for cases {\bf{A}} and {\bf{D}}.}
\\
- \tan(\alpha) |\sigma| &\textrm{for cases {\bf{B}} and {\bf{C}}.}
\end{array} \right .  
\ee
Thus, in cases {\bf{A}} and {\bf{D}}, the bion tilts in agreement 
with section \ref{sec-D3}.  In the other two cases,
it tilts in the opposite direction.  The interpretation
is that D1-branes coming `towards' the $D3$-brane
correspond to a positively charged magnetic monopole, 
while D1-branes coming `away from' the $D3$-brane
correspond to a negatively charged one.  Similarly, 
$\overline{\textrm{D1}}$-strings coming `away from' the $D3$-brane
correspond to a positively charged magnetic monopole, 
while D1-branes coming `towards' the $D3$-brane
correspond to a negatively charged one. The geometry of the
tilted bion inferred from the nonabelian dynamics of D1-branes
perfectly matches the results of the abelian D3-brane
calculation summarized in the previous section.

\section{Flat D$(p+r)$-brane from D$p$-branes}
\label{sec-plane} 
It is known that within Yang-Mills theory, lower
dimensional branes can expand to form higher
dimensional noncommutative branes (see, for example,
\cite{seiberg} and references therein).  In this section, we
show how this construction can be extended to the nonlinear
case of the nonabelian BI action. The point of this
exercise (which looks like a detour from the line of
argument of the rest of the paper) is to infer a specific
recipe for evaluating the worldvolume gauge fields in
the noncommutative description of a D-brane. In the
next section we will apply the spirit of this recipe
to the curved branes which are our primary interest.

We take the spacetime metric to be the flat Minkowski
metric ($g_{\mu\nu}=\eta_{\mu\nu}$), the dilaton to be constant
and the worldvolume gauge field to be zero. We take the 
background two-form field $B$ to have nonzero (constant) 
components only in directions transverse to the brane, 
$i,j,k = p+1,\ldots,9$. The world-volume of the branes
is parametrized by $X^a = \sigma^a, a=0,\ldots,p$ 
({\it{i.e.}} we are using the static gauge) and the transverse 
fluctuations are $X^i = \lambda \Phi^i$, where
$\Phi^i$ are $N\times N$ matrices in the adjoint of the
gauge group.  

We will look for solutions where the transverse scalars 
$\Phi$ are not functions of the brane coordinates $\sigma^a$.
In this case, the action (\ref{eqn:SBI}) for the nonabelian
dynamics of N Dp-branes reduces to
\be
\label{eqn:constact}
S = -g T_p \int d^p x^a
STr \left ( \sqrt{det(Q^i_j) }\right )~.
\ee
where the explicit form of $Q$ is displayed in (\ref{eqn:Q}). 
It is easy to to show that matrices $\Phi^i$, satisfying
\be
[\Phi^i, \Phi^j] = \frac{i} {\lambda^2}
\theta^{ij} I_{N\times N}~,
\label{eqn:soln}
\ee
where $\theta^{ij}$ is an arbitrary $(9-p)\times(9-p)$ 
antisymmetric matrix of c-numbers, solve the equations of 
motion.  Substituting this solution into the general 
definition of $Q$, (\ref{eqn:Q}),
gives
\be
\label{eqn:Qeval}
Q^i_j = \delta^i_j - \lambda^{-1} \theta^{il} (g+\lambda B)_{lj}~.
\ee  
$\theta$ can have any even rank $r$ up to $9-p$. With no loss of
generality, we can block diagonalize $\theta$ so that
$\theta^{\mu\nu} \neq 0$ for $\mu,\nu = p+1,\ldots,p+r$.
The remaining directions will be denoted by $m,n =
p+r+1,\ldots,9$ and of course $\theta^{i n} = 0$ for $i, n$ in
their appropriate ranges. From now on, $\theta=\theta^{\mu\nu}$
will denote an invertible, $r\times r$ matrix with inverse
$\theta_{\mu\nu}$.  Also, let us restrict our attention to
background two-form fields $B$ only in directions
$p+1,\ldots,p+r$, since the other components can simply be
gauged away.

The above can be summarized by saying that solution
(\ref{eqn:soln}) divides the $\Phi^i$'s into $r/2$ pairs 
satisfying canonical commutation relations and $9-p-r$ other
commuting coordinates (which we can drop from further consideration).
The slight hitch is that canonical commutation relations can 
only be realized on infinite-dimensional function spaces, and
not on finite-dimensional matrices. The solution only
makes sense if we take $N\to\infty$ and reinterpret all matrix 
operations (multiplication, trace, etc.) in the action as the 
corresponding Hilbert space operations. Fortunately, the 
technology for doing this sort of thing has been worked out in 
the study of noncommutative field theory over the past couple 
of years. Indeed, the physics of small fluctuations about 
(\ref{eqn:soln}) is best described by a noncommutative
field theory on the r-dimensional base space spanned by the
noncommuting $\Phi^i$. It is in this sense that we will
interpret the existence of $r$ noncommuting transverse 
displacement fields on the Dp-brane as creating
an effective D(p+r)-brane.  We will first show that the 
energetics of  (\ref{eqn:soln}) are indistinguishable from 
that of an abelian D(p+r)-brane with a certain worldvolume 
gauge field (which is the quantity that is the focus of our interest).

To establish the desired result, we make use of an equivalence
established in noncommutative field theory between actions
built on ordinary integrals of functions of ordinary coordinates, 
but with a noncommutative definition of multiplication of
functions (the $*$-product or Moyal product) and actions where 
functions become operators on a harmonic oscillator Hilbert 
space and the action is computed as the trace of an operator on
that Hilbert space (integral over space becomes Hilbert space trace) 
\cite{ncg,harvard}. In our interpretation of (\ref{eqn:soln}), the
$STr$ operation in the action (\ref{eqn:constact}) is to be 
thought of as a trace over operators on a Hilbert space. As just 
indicated, the trace can be recast as an integral over the
associated noncommuting coordinates, but we need the exact
relative normalization. The identification we need has been 
worked out in the noncommutative field theory literature \cite{seiberg}:
\be
\label{eqn:equiv}
\ d^r x^\mu \longleftrightarrow (2\pi)^{r \over 2} 
{\mathrm{Pf}}(\theta) STr~,
\ee
where ${\mathrm{Pf}}$ denotes the Pfaffian: 
${\mathrm{Pf}}(\theta)^2 = det(\theta)$. Putting the various pieces
of the puzzle together, we can express the action for our 
solution in the form of an equivalent integral of an energy density
over a D(r+p)-brane worldsheet:
\bear
S(\theta) &=& -g T_p \int d^p x^a d^r x^\mu
{1 \over \sqrt{2\pi det(\theta)}}
\left ( \sqrt{det(\delta^i_j - \lambda^{-1} \theta^{il} 
                (g+\lambda B)_{lj}) }\right )
\nn\\
&=& -g T_p (2 \pi \lambda)^{-{r \over 2}} \int d^p x^a d^r x^\mu
\left ( \sqrt{det(g+\lambda B - \lambda \theta^{-1}) }\right )~.
\eear
Since $T_{p+r} = T_p (2 \pi \lambda)^{-{r \over 2}}$,
the object we have constructed has the right energy
density to be a D(p+r)-brane with a world-volume F-flux
equal to $-\theta^{-1}$. Noncommutative coordinates for a 
lower-dimensional brane have, in a simple context, been converted
to a higher dimensional brane carrying a worldvolume gauge field.
For future reference, the interesting thing is the way the gauge field
arises via the commutator of the lower-dimensional matrix 
coordinates (\ref{eqn:soln}).

To give this idea a more demanding test, we will now check whether
we can reproduce the correct Chern-Simon couplings.  We will need
the nonabelian Chern-Simons action proposed by Myers in \cite{dielectric}
\be
\label{eqn:SCS}
S_{CS} = \mu_p \int STr \left ( P \left [
e^{i\lambda i_\Phi i_\Phi} (\Sigma C^{(n)} e^{\lambda B})
\right ] e^{\lambda {F}} \right )
\ee
(we refer the reader to Myers' paper for the definition of the symbol
$\lambda i_\Phi i_\Phi$ and other notation).
For concreteness, specialize to p=1 (nonabelian D-strings extending
in the 01-directions), with $\theta$ and $B$ nonzero only in the
23-directions.
Examine the coupling to $C^{(2)}$, specifically, to the $C_{01}$
component.  Expand the action (\ref{eqn:SCS}) and pick off the
coupling of interest to obtain
\be
S = \mu_1 \int dx^0 dx^1 STr C_{01} \left(
(1 + (i\lambda i_\Phi i_\Phi) \lambda B
\right )~.
\ee
Using the solution (\ref{eqn:soln}) for $\Phi$, we have
$(i\lambda i_\Phi i_\Phi) \lambda B = i\lambda^2[\Phi^1,\Phi^2] B_{12}
= -\theta B$. Passing from $STr$ to $\int$ according to (\ref{eqn:equiv})
we obtain an expression for this interaction in terms of an integral
over the D$(p+r)$-brane worldvolume:
\be
S = \mu_3 \int dx^0 dx^1 dx^2 dx^3 C_{01} \left(
B - \theta^{-1} \right ).
\ee
This is precisely the right coupling for a D3-brane with a world-volume
field $F_{23} = -\theta^{-1}$. The world-volume field $F$ suggested
here corresponds precisely to the field which one would expect to get
from the D-strings dissolved in a D3-brane with density $\theta^{-1}$
per area normal to the D-strings.  Solution (\ref{eqn:soln}) corresponds
to exactly this density of D-strings.

The lesson we learn from this computation is that when D$p$-branes
expand to form a D$(p+r)$-brane, the world volume gauge field F
on the D$(p+r)$-brane can be computed from the inverse of the density
of D$p$-branes, which in turn can be obtained from the commutators
of the transverse coordinates.

\section{The Worldvolume Gauge Field on the Dual Bion}
\label{sec-curved} 
In this section, we will take the prescription given in
section \ref{sec-plane} for identifying the worldvolume
gauge field implicit in a set of noncommuting coordinates
and adapt it to the bion problem under discussion.

We begin with the simple example of $N$ D1-branes ($N$ 
large) with $B=0$ (the setup of section \ref{sec-D1}). 
Choose the solution (\ref{eqn:ansatz0}) based on the 
$N\times N$ representation of $SU(2)$ and further
specialize to the representation where $\a^3$ is diagonal: 
$\a^3 = \mbox{diag}(N-1, N-3, N-5, \ldots, -N+3, -N+1)$.
The sphere described by equation (\ref{eqn:ansatz0}) at a 
fixed $\sigma$ goes through a point $(X^1, X^2, X^3) = 
(0, 0, R = \pi\a' N/|\sigma|)$.
A small patch of the sphere near this point is described
by the corner $k\times k$ blocks of the full $SU(2)$ representation
matrixes $\a^i$, where $k \ll N$.  Explicitly, replace
$\a^3$ with $\mbox{diag}(N-1, N-3, \ldots, N-2k+1)$, which for $k \ll N$ is 
approximately just
$N~I_{k \times k}$.  The small patch of the sphere is now
described by the same commutator as the noncommutative plane 
(section \ref{sec-plane}),
\be
[\Phi^1, \Phi^2] = i (2 \hat R) \Phi^3 \rightarrow 
\pm i {N \over 2 \sigma^2}~.
\ee
The `$+$'-sign corresponds to cases {\bf{B}} and {\bf{C}}, and
the `$-$'-sign corresponds to cases {\bf{A}} and {\bf{D}} in section \ref{sec-D1}.
Following section \ref{sec-plane}, we now write
\be
\theta^{12} = -i \lambda^2 [\Phi^1, \Phi^2] = {\lambda^2 N \over 2 \sigma^2}~,
\ee
so that the identification $F = -\theta^{-1}$ gives
\be
F_{12} = \pm {2 \sigma^2 \over \lambda^2 N} = \pm {N \over 2 R^2}~.
\ee
Referring back to (\ref{eqn:Ffield}) in section \ref{sec-D3}, we see that
this is indeed the correct value of the worldvolume gauge field.
Again, in cases {\bf{A}} and {\bf{D}}, we obtain the `$-$'-sign while in cases
{\bf{B}} and {\bf{C}}, we obtain the opposite sign and monopole charge.
This is all in agreement with expectations from (\ref{eqn:Ffield}).
The essence of this computation is that the commutator $[\Phi^i, \Phi^j]$
defines a two-form field in the worldvolume of the D3-brane, whose inverse
is the worldvolume gauge field $F$.

Computing the worldvolume gauge field in case of $B\neq0$ 
is very similar, except the geometry is more complicated.
To avoid any formulas with multiple $\pm$ signs, we will
specialize to case {\bf{A}} above, 
choosing $\sigma<0$ and $\hat R = (2\sigma)^{-1}$ 
(the other three cases are similar).
We want to evaluate the gauge field on the tilted
D3-brane implied by the nonabelian solution (\ref{eqn:ansatzB}) 
and compare it to the result of a direct calculation given in 
(\ref{eqn:Ffield}). To do this, it is best to convert 
(\ref{eqn:ansatzB}) to the coordinates given in (\ref{eqn:y-coord}). 
Defining rotated variables
\be
\Psi^1=\Phi^1~, \quad \Psi^2=\Phi^2~, 
\quad \Psi^3=\cos(\a) \Phi^3 - \sin(\a) \sigma/\lambda~, \quad
\Psi^4=\sin(\a) \Phi^3 + \cos(\a) \sigma/\lambda~
\ee
and then inserting (\ref{eqn:ansatzB}) gives
\bear
\label{eqn:rotcoord}
\Psi^1 &=& \cos(\a) \inv{2\sigma}~ \a^1 ~,\nn\\
\Psi^2 &=& \cos(\a) \inv{2\sigma}~ \a^2 ~, \\
\Psi^3 &=& \cos(\a) \inv{2\sigma}~ \a^3 ~,\nn\\
\Psi^4 &=& \sin(\a) \inv{2\sigma}~ \a^3 + 
                \inv{\cos(\a)}~ {\sigma\over\lambda} ~.\nn
\eear
To check that this makes sense, take $N$ large and pass to 
the classical limit, by setting $\a^i \rightarrow N n^i$ 
where $(n^1)^2+(n^2)^2+(n^3)^2 = 1$.  The
$\Psi$'s become classical coordinates
\bear
\lambda \Psi^1 &\rightarrow Y^1 =&\cos(\a) {N\pi\a' \over \sigma} ~n^1~, \nn\\
\lambda \Psi^2 &\rightarrow Y^2 =&\cos(\a) {N\pi\a' \over \sigma} ~n^2~, \\
\lambda \Psi^3 &\rightarrow Y^3 =&\cos(\a) {N\pi\a' \over \sigma} ~n^3~, \nn\\
\lambda \Psi^4 &\rightarrow Y^4 =&\sin(\a) {N\pi\a' \over \sigma} ~n^3 
+ \inv{\cos(\a)}~ \sigma~ \nn\\
        &   =&\tan(\a)Y_3 + {N\pi\a'\over\sqrt{Y_1^2+Y_2^2+Y_3^3}}~,\nn
\eear
in perfect correspondence with equation (\ref{eqn:bionY}).

In section \ref{sec-plane} we showed that the worldvolume gauge field
is computed from the commutators of the transverse scalars. Using
(\ref{eqn:rotcoord}) to compute the commutators and comparing with 
(\ref{eqn:soln}), we obtain the following noncommutativity tensor $\Theta$:
\bear
-i\lambda^2[\Psi^i, \Psi^j] = 
        2 \e^{ijk} {\lambda^2 \cos(\a) \over 2\sigma} \Psi^k 
&\rightarrow& { \lambda \cos(\a) \over\sigma} \e^{ijk} Y^k 
        \equiv \Theta^{ij}~, \nn\\
-i\lambda^2[\Psi^1, \Psi^4] = - 2 {\lambda^2 \sin(\a) \over 2\sigma} \Psi^2 
&\rightarrow& - {\lambda \sin(\a) \over \sigma} Y^2 \equiv \Theta^{14}~, \\
-i\lambda^2[\Psi^2, \Psi^4] = 2 {\lambda^2 \sin(\a) \over 2\sigma} \Psi^1 
&\rightarrow & {\lambda \sin(\a) \over \sigma} Y^1 \equiv \Theta^{24}~, \nn\\
-i\lambda^2[\Psi^3, \Psi^4] = 0 &\rightarrow& 0 \equiv \Theta^{34}~, \nn
\eear
where $i,j,k=1,\ldots,3.$
Finally, we need to pull the two-tensor $\Theta$ back to the worldvolume
of the D3-brane, to the worldvolume coordinates of (\ref{eqn:bionY}).  
With a little algebra, it can be checked that 
$\Theta^{\mu\nu}$ ($\mu,\nu =1,\dots,4$) satisfies
\be
\Theta^{\mu\nu} = 
\frac{\partial Y^\mu}{\partial y^i}
\frac{\partial Y^\nu}{\partial y^j}
\theta^{ij}~,
\label{eqn:miracle}
\ee
where
\be
\theta^{ij} = { \lambda \cos(\a) \over \sigma} ~\e^{ijk} y^k = 
- {2 \sqrt{(y^1)^2+(y^2)^2+(y^3)^2} \over N} ~\e^{ijk} y^k ~
\label{eqn:thetensor}
\ee
(the minus sign is a consequence of our having chosen case {\bf{A}}, $\sigma< 0$).
According to section \ref{sec-plane}, the worldvolume gauge field is
the negative inverse of the noncommutativity tensor $\theta^{ij}$.
However, the tensor of (\ref{eqn:thetensor}) is not invertible: it 
acts in three dimensions and has one zero eigenvalue. Following section 
\ref{sec-plane}, the inversion of $\theta^{ij}$ is to be carried out on 
the subspace orthogonal to the subspace of zero eigenvalues. With 
this understanding, we obtain the following result for the gauge 
field on the D3-brane
\be
F_{ij}= (-\theta^{-1})_{ij} = -{N \over 2 [(y^1)^2+(y^2)^2+(y^3)^2]^{3/2}}
~\e^{ijk} y^k ~,
\ee
in perfect agreement with the abelian D3-brane result,
equation (\ref{eqn:Ffield}). This is what we wanted to show.

\section{Conclusion}
We have shown that the noncommutative bion solution 
from \cite{ncbion} can be generalized to include 
a nonzero NS-NS two-form field $B$.  The geometry 
extracted from our generalized solution agrees 
with the `dual' picture provided by the abelian 
theory on a D3-brane in the presence of a nonzero 
$B$.  Even better, we have been able to argue that the
nonabelian calculation makes a prediction for the worldvolume
gauge field on the D3-brane and we find that this agrees
with the abelian caluculation as well. Although limited
to BPS configurations, we regard these considerations
as a significant further test of the validity of 
the nonabelian Born-Infeld action 
proposed by Myers in \cite{dielectric}.


\section*{Acknowledgments}
The research of JLK is supported in part by the Natural Sciences 
and Engineering Research Council of Canada. The research of
CGC is supported in part by US Department of Energy grant
number DE-FG02-91ER40671. CGC also wishes to thank the Laboratoire
de Physique Th\'eoriqe de l'\'Ecole Normale Sup\'erieure for
hospitality during the concluding phases of the work reported here.


\def\np#1#2#3{{\it Nucl.\ Phys.} {\bf B#1} (#2) #3}
\def\pl#1#2#3{{\it Phys.\ Lett.} {\bf B#1} (#2) #3}
\def\physrev#1#2#3{{\it Phys.\ Rev.\ Lett.} {\bf #1} (#2) #3}
\def\prd#1#2#3{{\it Phys.\ Rev.} {\bf D#1} (#2) #3}
\def\ap#1#2#3{{\it Ann.\ Phys.} {\bf #1} (#2) #3}
\def\ppt#1#2#3{{\it Phys.\ Rep.} {\bf #1} (#2) #3}
\def\rmp#1#2#3{{\it Rev.\ Mod.\ Phys.} {\bf #1} (#2) #3}
\def\cmp#1#2#3{{\it Comm.\ Math.\ Phys.} {\bf #1} (#2) #3}
\def\mpla#1#2#3{{\it Mod.\ Phys.\ Lett.} {\bf #1} (#2) #3}
\def\jhep#1#2#3{{\it JHEP} {\bf #1} (#2) #3}
\def\atmp#1#2#3{{\it Adv.\ Theor.\ Math.\ Phys.} {\bf #1} (#2) #3}
\def\jgp#1#2#3{{\it J.\ Geom.\ Phys.} {\bf #1} (#2) #3}
\def\cqg#1#2#3{{\it Class.\ Quant.\ Grav.} {\bf #1} (#2) #3}

\def\hepth#1{{\it hep-th/{#1}}}





\begin{thebibliography}{99}

\bibitem {dielectric} {R.C.Myers, {``Dielectric-branes,''}
\jhep {9912} {1999} {022}, {\tt hep-th/9910053}.}

\bibitem {symtrace} {A. A. Tseytlin, {``On non-abelian generalisation 
of Born-Infeld action in string theory,''}
\np {B501} {1997} {41}, {\tt hep-th/9701125}.}

\bibitem{mark} {W. Taylor, M. Van Raamsdonk,
{``Multiple D0-branes in Weakly Curved Backgrounds,''}
\np {B558} {1999} {63}, {\tt hep-th/9904095};}
{W. Taylor, M. Van Raamsdonk,
{``Multiple Dp-branes in Weak Background Fields,''}
\np {B573} {2000} {703}, {\tt hep-th/9910052}.}

\bibitem {ncbion} {N.R.Constable, R.C.Myers, O.Tafjord
{``The Noncommutative Bion Core,''} 
\prd {61} {2000} {106009}, {\tt hepth/9911136}.}
\bibitem {callan} {C.G.Callan, J.M.Maldacena,
{``Brane Dynamics From the Born-Infeld Action,''}
\np {513} {1998} {198}, {\tt hep-th/9708147}.}
\bibitem {gibbons} {G.W.Gibbons, 
{``Born-Infeld particles and Dirichlet p-branes''} 
\np{514}{1998}{603}, {\tt hep-th/9709027}.}

\bibitem {seiberg} {N.Seiberg,
{``A Note on Background Independence in Noncommutative Gauge Theories,
Matrix Model and Tachyon Condensation,''}
\jhep {0009} {2000} {003}, {\tt hep-th/0008013}.}

\bibitem {sw} {N.Seiberg, E.Witten 
{``String Theory and Noncommutative Geometry,''} 
\jhep{9909} {1999} {032}, {\tt hep-th/9908142}.}

\bibitem {hashimoto} {K. Hashimoto, {``Born-Infeld
Dynamics in Uniform Magnetic Field,''}
\jhep{9907} {1999} {016}, {\tt hep-th/9905162}.}
\bibitem {moriyama} {S. Moriyama, {``Noncommutative
Monopole from Nonlinear Monopole,''} 
\pl{485} {2000} {278}, {\tt hep-th/0003231}.}

\bibitem {nahm0} {D. E. Diaconescu, {``D-branes, Monopoles
and Nahm Equations,''} 
\np{503} {1997} {220}, {\tt hep-th/9608163}.}

\bibitem {nahmB} {D.Bak, {``Deformed Nahm Equation and a 
Noncommutative BPS Monopole,''} 
\pl {471} {1999} {149}, {\tt hep-th/9910135}.}

\bibitem{ncg} {A. Connes, M. R. Douglas, A. Schwarz,
{``Noncommutative Geometry and Matrix Theory: Compactification on Tori,''}
\jhep{9802}{1998}{003}, {\tt hep-th/9711162}.}

\bibitem{harvard} {R. Gopakumar, S. Minwalla, A. Strominger,
{``Noncommutative Solitons,''} \jhep{0005} {2000} {020},
{\tt hep-th/0003160}.}



  
\end{thebibliography}
\end{document}